\pdfoutput=1
\documentclass[12pt,a4paper]{scrartcl}

\usepackage{ifthen} 
\newboolean{pdflatex}
\setboolean{pdflatex}{true} 

\newboolean{articletitles}
\setboolean{articletitles}{true} 

\newboolean{uprightparticles}
\setboolean{uprightparticles}{false} 

\def\paperasciititle{Lepton-flavor violation and lepton-flavor-universality violation in b and c decays} 
\def\papertitle{Lepton-flavor violation and lepton-flavor-universality violation\\in \bquark and \cquark decays} 
\def\paperkeywords{{High Energy Physics}, {LHCb}} 
\def\papercopyright{}

\def\paperlicenceurl{https://creativecommons.org/licenses/by/4.0}

\usepackage[top=1in, bottom=1.25in, left=1in, right=1in]{geometry}

\usepackage{microtype}
\usepackage{lineno}  
\usepackage{xspace} 
\usepackage{caption} 

\usepackage{graphicx}  
\usepackage{color}
\usepackage{colortbl}
\graphicspath{{./figs}} 

\usepackage{amsmath} 
\usepackage{amssymb}
\usepackage{amsfonts}
\usepackage{upgreek} 

\newcommand*\patchAmsMathEnvironmentForLineno[1]{%
\expandafter\let\csname old#1\expandafter\endcsname\csname #1\endcsname
\expandafter\let\csname oldend#1\expandafter\endcsname\csname
end#1\endcsname
 \renewenvironment{#1}%
   {\linenomath\csname old#1\endcsname}%
   {\csname oldend#1\endcsname\endlinenomath}%
}
\newcommand*\patchBothAmsMathEnvironmentsForLineno[1]{%
  \patchAmsMathEnvironmentForLineno{#1}%
  \patchAmsMathEnvironmentForLineno{#1*}%
}
\AtBeginDocument{%
\patchBothAmsMathEnvironmentsForLineno{equation}%
\patchBothAmsMathEnvironmentsForLineno{align}%
\patchBothAmsMathEnvironmentsForLineno{flalign}%
\patchBothAmsMathEnvironmentsForLineno{alignat}%
\patchBothAmsMathEnvironmentsForLineno{gather}%
\patchBothAmsMathEnvironmentsForLineno{multline}%
\patchBothAmsMathEnvironmentsForLineno{eqnarray}%
}

\usepackage{hyperxmp}

\usepackage[pdftex,
            pdfauthor={Koppenburg,Hiller},
            pdftitle={\paperasciititle},
            pdfkeywords={\paperkeywords},
            pdfcopyright={Copyright (C) \papercopyright},
            pdflicenseurl={\paperlicenceurl}]{hyperref}

\usepackage[colorinlistoftodos,textsize=scriptsize]{todonotes}

\usepackage[bottom,flushmargin,hang,multiple]{footmisc}

\usepackage[all]{hypcap} 

\usepackage{xspace} 
\usepackage{upgreek}

\def\lhcb   {\mbox{LHCb}\xspace}

\def\MagUp {\mbox{\em Mag\kern -0.05em Up}\xspace}

\ifthenelse{\boolean{uprightparticles}}%
{

 \def\Pmu         {\ensuremath{\upmu}\xspace}                 
 \def\Pnu         {\ensuremath{\upnu}\xspace}                 
                  
 \def\Ppi         {\ensuremath{\uppi}\xspace}

 \def\Ptau        {\ensuremath{\uptau}\xspace}                 
                  
 \def\Pphi        {\ensuremath{\upphi}\xspace}

 \def\Ppsi        {\ensuremath{\uppsi}\xspace}

 \def\PDelta      {\ensuremath{\Delta}\xspace}                 
 \def\PXi         {\ensuremath{\Xi}\xspace}                 
 \def\PLambda     {\ensuremath{\Lambda}\xspace}                 
 \def\PSigma      {\ensuremath{\Sigma}\xspace}                 
 \def\POmega      {\ensuremath{\Omega}\xspace}                 
 \def\PUpsilon    {\ensuremath{\Upsilon}\xspace}

 \def\PB      {\ensuremath{\mathrm{B}}\xspace}                 
                  
 \def\PD      {\ensuremath{\mathrm{D}}\xspace}

 \def\PJ      {\ensuremath{\mathrm{J}}\xspace}                 
 \def\PK      {\ensuremath{\mathrm{K}}\xspace}

 \def\PZ      {\ensuremath{\mathrm{Z}}\xspace}                 
                  
 \def\Pb      {\ensuremath{\mathrm{b}}\xspace}                 
 \def\Pc      {\ensuremath{\mathrm{c}}\xspace}                 
                  
 \def\Pe      {\ensuremath{\mathrm{e}}\xspace}

 \def\Pi      {\ensuremath{\mathrm{i}}\xspace}

 \def\Ps      {\ensuremath{\mathrm{s}}\xspace}

 \def\thebaroffset{0.0em}
}
{

 \def\Pmu         {\ensuremath{\mu}\xspace}                 
 \def\Pnu         {\ensuremath{\nu}\xspace}                 
                  
 \def\Ppi         {\ensuremath{\pi}\xspace}

 \def\Ptau        {\ensuremath{\tau}\xspace}                 
                  
 \def\Pphi        {\ensuremath{\phi}\xspace}

 \def\Ppsi        {\ensuremath{\psi}\xspace}                 
                  
 \mathchardef\PDelta="7101
 \mathchardef\PXi="7104
 \mathchardef\PLambda="7103
 \mathchardef\PSigma="7106
 \mathchardef\POmega="710A
 \mathchardef\PUpsilon="7107
                  
 \def\PB      {\ensuremath{B}\xspace}                 
                  
 \def\PD      {\ensuremath{D}\xspace}

 \def\PJ      {\ensuremath{J}\xspace}                 
 \def\PK      {\ensuremath{K}\xspace}

 \def\PZ      {\ensuremath{Z}\xspace}                 
                  
 \def\Pb      {\ensuremath{b}\xspace}                 
 \def\Pc      {\ensuremath{c}\xspace}                 
                  
 \def\Pe      {\ensuremath{e}\xspace}

 \def\Pi      {\ensuremath{i}\xspace}

 \def\Ps      {\ensuremath{s}\xspace}

 \def\thebaroffset{0.18em}
}
\newcommand{\offsetoverline}[2][\thebaroffset]{\kern #1\overline{\kern -#1 #2}}%

\makeatletter
\ifcase \@ptsize \relax
  \newcommand{\miniscule}{\@setfontsize\miniscule{4}{5}}
\or
  \newcommand{\miniscule}{\@setfontsize\miniscule{5}{6}}
\or
  \newcommand{\miniscule}{\@setfontsize\miniscule{5}{6}}
\fi
\makeatother

\DeclareRobustCommand{\optbar}[1]{\shortstack{{\miniscule (\rule[.5ex]{1.25em}{.18mm})}
  \\ [-.7ex] $#1$}}

\def\ep         {{\ensuremath{\Pe^+}}\xspace}
\def\epm        {{\ensuremath{\Pe^\pm}}\xspace} 
 
\def\epem       {{\ensuremath{\Pe^+\Pe^-}}\xspace}

\def\mup        {{\ensuremath{\Pmu^+}}\xspace}
 
\def\mump       {{\ensuremath{\Pmu^\mp}}\xspace} 
\def\mumu       {{\ensuremath{\Pmu^+\Pmu^-}}\xspace}

\def\taup       {{\ensuremath{\Ptau^+}}\xspace}
\def\taum       {{\ensuremath{\Ptau^-}}\xspace}
\def\taupm      {{\ensuremath{\Ptau^\pm}}\xspace}

\def\ellm       {{\ensuremath{\ell^-}}\xspace}
\def\ellp       {{\ensuremath{\ell^+}}\xspace}

\def\ellell     {\ensuremath{\ell^+ \ell^-}\xspace}

\def\neu        {{\ensuremath{\Pnu}}\xspace}
\def\neub       {{\ensuremath{\overline{\Pnu}}}\xspace}

\def\Z      {{\ensuremath{\PZ}}\xspace}

\def\squark    {{\ensuremath{\Ps}}\xspace}

\def\cquark    {{\ensuremath{\Pc}}\xspace}
\def\cquarkbar {{\ensuremath{\overline \cquark}}\xspace}

\def\bquark    {{\ensuremath{\Pb}}\xspace}
\def\bquarkbar {{\ensuremath{\overline \bquark}}\xspace}

\def\pion   {{\ensuremath{\Ppi}}\xspace}

\def\pim    {{\ensuremath{\pion^-}}\xspace}

\def\kaon    {{\ensuremath{\PK}}\xspace}
\def\Kbar    {{\ensuremath{\offsetoverline{\PK}}}\xspace}
\def\Kb      {{\ensuremath{\Kbar}}\xspace}
\def\KorKbar {\kern \thebaroffset\optbar{\kern -\thebaroffset \PK}{}\xspace}

\def\Kp      {{\ensuremath{\kaon^+}}\xspace}

\def\Kpm     {{\ensuremath{\kaon^\pm}}\xspace}

\def\KS      {{\ensuremath{\kaon^0_{\mathrm{S}}}}\xspace}

\def\Kstarz  {{\ensuremath{\kaon^{*0}}}\xspace}

\def\Kstar   {{\ensuremath{\kaon^*}}\xspace}

\def\D       {{\ensuremath{\PD}}\xspace}

\def\DorDbar {\kern \thebaroffset\optbar{\kern -\thebaroffset \PD}\xspace}

\def\Dp      {{\ensuremath{\D^+}}\xspace}
\def\Dm      {{\ensuremath{\D^-}}\xspace}

\def\DpDm    {\ensuremath{\Dp {\kern -0.16em \Dm}}\xspace}
\def\Dstar   {{\ensuremath{\D^*}}\xspace}

\def\Ds      {{\ensuremath{\D^+_\squark}}\xspace}

\def\B       {{\ensuremath{\PB}}\xspace}
\def\Bbar    {{\ensuremath{\offsetoverline{\PB}}}\xspace}
\def\Bb      {{\ensuremath{\Bbar}}\xspace}
\def\BorBbar {\kern \thebaroffset\optbar{\kern -\thebaroffset \PB}\xspace}
\def\Bz      {{\ensuremath{\B^0}}\xspace}

\def\Bd      {{\ensuremath{\B^0}}\xspace}

\def\BdorBdbar {\kern \thebaroffset\optbar{\kern -\thebaroffset \Bd}\xspace}
\def\Bu      {{\ensuremath{\B^+}}\xspace}

\def\Bp      {{\ensuremath{\Bu}}\xspace}

\def\Bs      {{\ensuremath{\B^0_\squark}}\xspace}

\def\BsorBsbar {\kern \thebaroffset\optbar{\kern -\thebaroffset \Bs}\xspace}
\def\Bc      {{\ensuremath{\B_\cquark^+}}\xspace}

\def\Bds     {{\ensuremath{\B_{(\squark)}^0}}\xspace}

\def\BdorBs  {\Bds}

\def\jpsi     {{\ensuremath{{\PJ\mskip -3mu/\mskip -2mu\Ppsi}}}\xspace}
\def\psitwos  {{\ensuremath{\Ppsi{(2S)}}}\xspace}

\def\Y#1S{\ensuremath{\PUpsilon{(#1S)}}\xspace}

\def\FourS {{\Y4S}\xspace}

\def\Lz          {{\ensuremath{\PLambda}}\xspace}

\def\LorLbar     {\kern \thebaroffset\optbar{\kern -\thebaroffset \PLambda}\xspace}

\def\Lc          {{\ensuremath{\Lz^+_\cquark}}\xspace}

\def\Lb           {{\ensuremath{\Lz^0_\bquark}}\xspace}

\def\BF         {{\ensuremath{\mathcal{B}}}\xspace}

\newcommand{\decay}[2]{\ensuremath{#1\!\to #2}\xspace} 

\def\to                 {\ensuremath{\rightarrow}\xspace}

\def\qsq       {{\ensuremath{q^2}}\xspace}

\def\eps   {{\ensuremath{\varepsilon}}\xspace}

\def\AT#1     {\ensuremath{A_{\mathrm{T}}^{#1}}\xspace}           

\def\C#1      {\ensuremath{\mathcal{C}_{#1}}\xspace}                       
\def\Cp#1     {\ensuremath{\mathcal{C}_{#1}^{'}}\xspace}                    
\def\Ceff#1   {\ensuremath{\mathcal{C}_{#1}^{\mathrm{(eff)}}}\xspace}        
\def\Cpeff#1  {\ensuremath{\mathcal{C}_{#1}^{'\mathrm{(eff)}}}\xspace}       
\def\Ope#1    {\ensuremath{\mathcal{O}_{#1}}\xspace}                       
\def\Opep#1   {\ensuremath{\mathcal{O}_{#1}^{'}}\xspace}                    

\newcommand{\nospaceunit}[1]{\ensuremath{\text{#1}}}       
\newcommand{\aunit}[1]{\ensuremath{\text{\,#1}}}       

\newcommand{\tev}{\aunit{Te\kern -0.1em V}\xspace}
\newcommand{\gev}{\aunit{Ge\kern -0.1em V}\xspace}
\newcommand{\mev}{\aunit{Me\kern -0.1em V}\xspace}
\newcommand{\kev}{\aunit{ke\kern -0.1em V}\xspace}
\newcommand{\ev}{\aunit{e\kern -0.1em V}\xspace}
 
\newcommand{\mevc}{\ensuremath{\aunit{Me\kern -0.1em V\!/}c}\xspace}
\newcommand{\gevc}{\ensuremath{\aunit{Ge\kern -0.1em V\!/}c}\xspace}
\newcommand{\mevcc}{\ensuremath{\aunit{Me\kern -0.1em V\!/}c^2}\xspace}
\newcommand{\gevcc}{\ensuremath{\aunit{Ge\kern -0.1em V\!/}c^2}\xspace}
\newcommand{\gevgevcccc}{\ensuremath{\gev^2\!/c^4}\xspace} 

\def\cm   {\aunit{cm}\xspace}

\def\mub{\ensuremath{\,\upmu\nospaceunit{b}}\xspace}
\def\nb {\aunit{nb}\xspace}

\def\fb   {\ensuremath{\aunit{fb}}\xspace}
\def\invfb   {\ensuremath{\fb^{-1}}\xspace}
\def\ab   {\ensuremath{\aunit{ab}}\xspace}
\def\invab   {\ensuremath{\ab^{-1}}\xspace}

\def\sec  {\ensuremath{\aunit{s}}\xspace}

\def\gsim{{~\raise.15em\hbox{$>$}\kern-.85em
          \lower.35em\hbox{$\sim$}~}\xspace}
\def\lsim{{~\raise.15em\hbox{$<$}\kern-.85em
          \lower.35em\hbox{$\sim$}~}\xspace}

\def\pt         {\ensuremath{p_{\mathrm{T}}}\xspace}

\def\tell1  {TELL1\xspace}
\def\ukl1   {UKL1\xspace}

\newcommand{\eg}{\mbox{\itshape e.g.}\xspace}
\newcommand{\ie}{\mbox{\itshape i.e.}\xspace}

\newcommand{\etc}{\mbox{\itshape etc.}\xspace}

\usepackage{cite} 
\usepackage{mciteplus}

\graphicspath{{figs/}} 

\usepackage{blindtext}
\usepackage{dcolumn}
\newcolumntype{M}{D{.}{\:}{3}}
\newcolumntype{d}[1]{D{.}{.}{#1}}
\newcolumntype{C}{>{$}c<{$}}
\newcolumntype{R}{>{$}r<{$}}
\newcolumntype{L}{>{$}l<{$}}

\newcommand{\aerr}[2]{{\:}^{+{\:}#1}_{-{\:}#2}}%
\usepackage{wrapfig}

\usepackage{color}
\definecolor{Cerulean}{rgb}{0.0, 0.48, 0.65}

\newcommand{\be}{\begin{equation}}
\newcommand{\ee}{\end{equation}}
\newcommand{\bea}{\begin{eqnarray}}
\newcommand{\eea}{\end{eqnarray}}

\newcommand{\mc}{\mathcal}

\newcommand{\noi}{\noindent}

\newcommand{\ale}{\alpha_{\rm e}}
\newcommand{\LaQCD}{\Lambda_{\rm QCD}}

\newcommand{\RD}{{\rm rd}}

\def\parenbar#1{{\null\!      
   \mathop#1\limits^{\hbox{%
   \tiny{{\fontsize{3.5pt}{0em}\selectfont (}%
   \raisebox{-0.4pt}{--}%
   {\fontsize{3.5pt}{0em}\selectfont )}}}} 
   \!\null}} 

\newcommand\snowmass{\begin{center}\rule[-0.2in]{\textwidth}{0.01in}\\\rule{\textwidth}{0.01in}\\
\vskip 0.1in Submitted to the Proceedings of the US Community Study\\ 
on the Future of Particle Physics (Snowmass 2021)\\ 
\rule{\textwidth}{0.01in}\\\rule[+0.2in]{\textwidth}{0.01in} \end{center}}

\DeclareOldFontCommand{\rm}{\normalfont\rmfamily}{\mathrm}
\DeclareOldFontCommand{\sf}{\normalfont\sffamily}{\mathsf}
\DeclareOldFontCommand{\tt}{\normalfont\ttfamily}{\mathtt}
\DeclareOldFontCommand{\bf}{\normalfont\bfseries}{\mathbf}
\DeclareOldFontCommand{\it}{\normalfont\itshape}{\mathit}
\DeclareOldFontCommand{\sl}{\normalfont\slshape}{\@nomath\sl}
\DeclareOldFontCommand{\sc}{\normalfont\scshape}{\@nomath\sc}

\begin{document}

\renewcommand{\thefootnote}{\fnsymbol{footnote}}
\setcounter{footnote}{1}

\begin{titlepage}

\snowmass
\vspace*{1.5cm}

{\bf\boldmath\huge
\begin{center}
\sffamily \papertitle
\end{center}
}

\vspace*{0.5cm}

\begin{center}
D.~Guadagnoli$^{1}$,
P.~Koppenburg$^{2}$
\bigskip\\
{\it\footnotesize
$^1$LAPTh, Universit\'{e} Savoie Mont-Blanc and CNRS, Annecy, France\\
$^2$Nikhef National Institute for Subatomic Physics, Amsterdam, Netherlands}
\end{center}

\vspace{\fill}

\begin{abstract}
\noindent Two topics have recently risen to prominence within the ongoing searches of beyond-Standard Model effects in $b$ and $c$ decays: observables that test lepton flavor universality (LFU) as well as lepton flavor violation (LFV). A coherent set of measurements suggests non-standard LFU effects. General arguments relate LFU to LFV, and the observed size of the former gives hope of observable signals for the latter. We attempt a comprehensive discussion of both theoretical and experimental aspects of these tests. The main final message is that all the instruments necessary to fully establish the putative new effects are at hand, thanks to running experiments and their upgrades. Therefore this subject stands concrete chances to usher genuinely unexpected discoveries.
\end{abstract}

\vspace{\fill}

\end{titlepage}

\pagestyle{empty}  


\renewcommand{\thefootnote}{\arabic{footnote}}
\setcounter{footnote}{0}

\tableofcontents
\cleardoublepage


\pagestyle{plain} 
\setcounter{page}{1}
\pagenumbering{arabic}


\section{Executive summary}

\begin{wrapfigure}{r}{0.55\textwidth}
\hskip 0.04\textwidth
\begin{minipage}{0.5\textwidth}
\includegraphics[width=\textwidth]{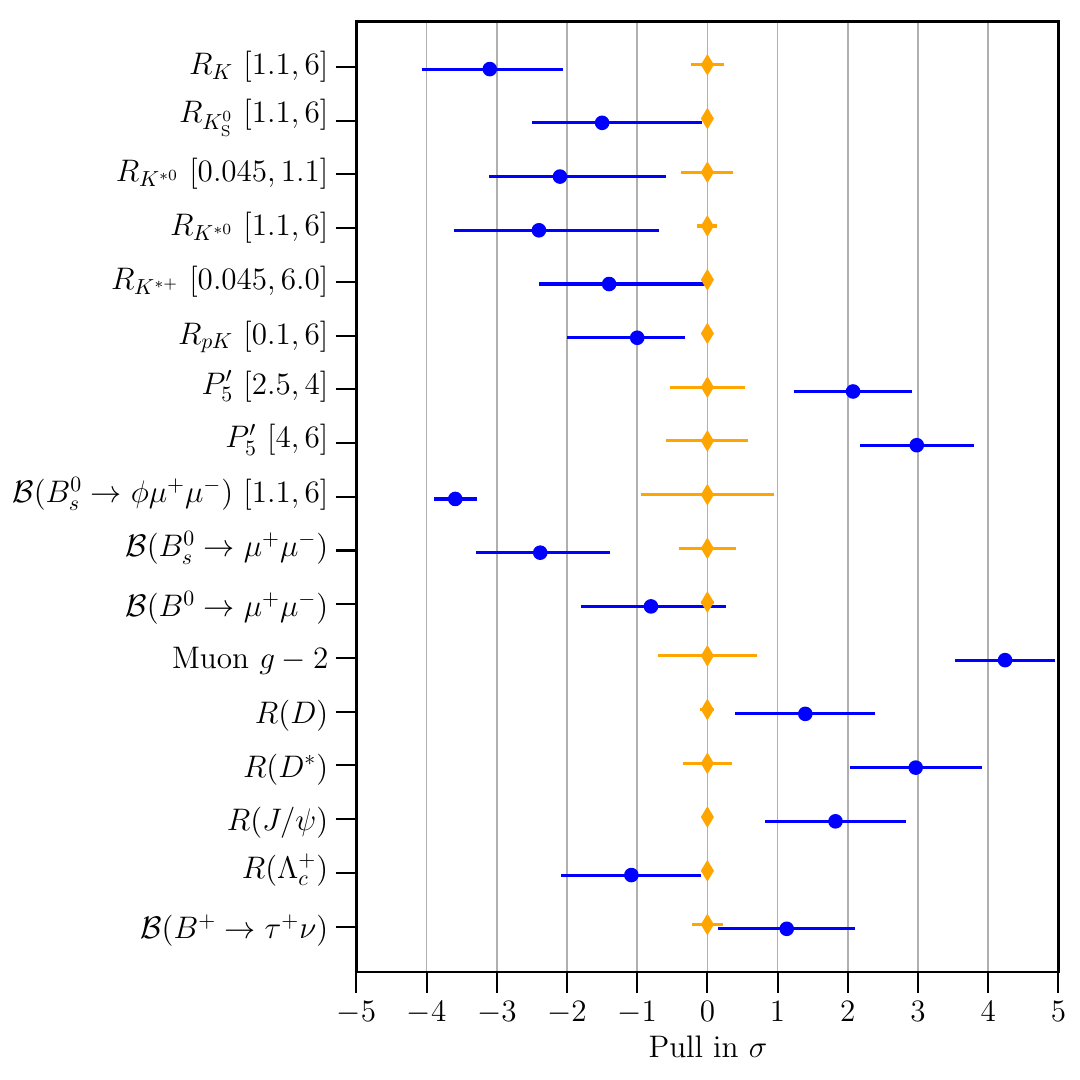}%
\caption{Selection of flavor anomalies shown as pulls of (blue points) experiment versus (orange diamonds) theory expectation~\cite{bllsPlots}. For each measurement the quadratic sum of the experimental and theory uncertainties is normalised to unity and the deviation of the experimental value is displayed in this unit. }\label{fig:Anomalies}
  \end{minipage}
\end{wrapfigure}
Results from the \B factories and the LHCb experiments show indications of deviations from the Standard Model in decays of bottom and charm hadrons. A selection of the most discrepant measurements is shown in Fig.~\ref{fig:Anomalies}. Many of these anomalies indicate deviations from lepton universality, a symmetry of the gauge sector --- and an {\em accidental} near-symmetry of the Yukawa sector --- of the Standard Model by which all leptons couple with the same strength. If confirmed, the observed deviations from lepton universality would, collectively, represent an unambiguous sign of New Physics.

Several ongoing and future experiments propose to further test these deviations with much larger data sets and improved detection and analysis strategies, improving both the statistical and the systematic uncertainties of the current measurements. Indeed, a large fraction of the measurements listed in this specific document have an experimental uncertainty much larger than the corresponding theoretical uncertainty on the SM prediction. This is likely to still be the case for the foreseeable future --- and this is why these measurements are known to be ``theoretically clean'' probes of New Physics.

Processes violating lepton flavor conservation have not been observed yet. The general expectation is however that they should be within experimental reach, if non-standard lepton universality violation is as large as measured. Since lepton flavor violation (LFV) is a null test of the SM, any measurement would be proof of New Physics. The very same experiments mentioned above also offer a broad program of LFV tests --- through a stream of analyses closely related to those aimed at lepton-universality tests.

\clearpage 
\section{Physics potential and reach}

The present document covers two families of processes: 
\begin{enumerate}
    \item Tests of lepton universality in semi-leptonic processes, where the SM predicts near-identical rates for all decays to light leptons, up to small radiative corrections well below the present and projected experimental accuracies.
    \item Searches for processes violating lepton flavor conservation.
\end{enumerate}
Within the SM, gauge interactions are lepton-universal; so-called Yukawa interactions, involving fermions and the Higgs field, are not lepton-universal because fermion masses are different from each other. This however implies power-suppressed lepton non-universal effects, that are tiny in decays to light leptons, and radiative corrections with lepton-mass-dependent logs that, again for light leptons, are ultimately below projected experimental sensitivities.
Further lepton-universality tests --- branching ratios and angular analyses --- have uncertainties dominated by the determination of the relevant hadronic matrix elements. Here, the comparison with experimental uncertainties has to be discussed case by case.
Finally, the SM predicts no lepton-flavor violation (LFV). Therefore, signals in item 2 are unequivocal proof of new phenomena.

The two above families of processes are, in general, intimately connected. In particular, the observed size of lepton-universality violation may lead to observable LFV as well.

The experimental reach in these processes will be dominated by the LHCb and Belle~II experiments, with additions from ATLAS, CMS and charm factories, described in more detail in Sec.~\ref{sec:experiments}.


\subsection{Most promising directions}\label{sec:most_promising}
At present there are hints of deviations from the SM in lepton-universality ratios,
which make them high-priority areas for further investigations. 

\begin{figure}[tb]
\centering
\includegraphics[width=0.7\textwidth]{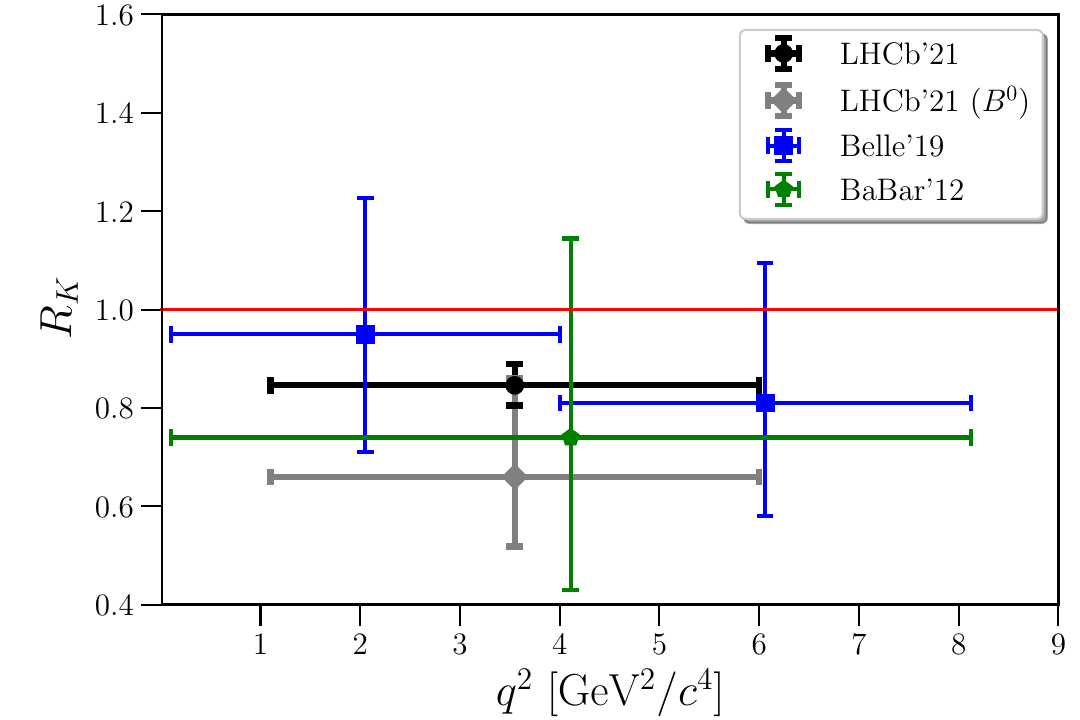}
\caption{Measurements of $R_K$ in the low-\qsq range~\cite{BaBar:2012mrf,BELLE:2019xld,LHCb:2021trn}.}\label{fig:RK}
\end{figure}

First are ratios in \decay{b}{s\ellell} decays defined as~\cite{Hiller:2003js}
\begin{equation}\label{eq:RX}
    R_X = \frac{%
            \int\limits_{q_\text{min}^2}^{q_\text{max}^2}{\rm d}\qsq\frac{d\Gamma\left(\decay{\B}{X\mumu}\right)}{{\rm d}\qsq}%
          }{%
            \int\limits_{q_\text{min}^2}^{q_\text{max}^2}{\rm d}\qsq\frac{d\Gamma\left(\decay{\B}{X\epem}\right)}{{\rm d}\qsq}}
\end{equation}
with $q^2=m^2(\ellell)$ and $X$ being a light hadron such as \Kpm, \KS, \Kstarz, \Pphi\etc The most precisely measured such ratio is $R_K$ using \decay{\Bp}{\Kp\ellell} decays. LHCb find $R_K=0.846\aerr{0.042}{0.039}\aerr{0.013}{0.012}$~\cite{LHCb:2021trn} for dileptons in the $1.1$--$6.0\gevgevcccc$ range, which is $3.1$ standard deviations below the expected value of unity. Similar --- though less significant --- deviations from unity are found in $R_\Kstar$~\cite{LHCb-PAPER-2017-013,LHCb-PAPER-2021-038}, or 
$R_{pK}$~\cite{LHCb-PAPER-2019-040} using \Lb baryons. 
The ratios $R_K$ and $R_{K^*}$ have also been measured at BaBar~\cite{BaBar:2012mrf} and Belle~\cite{Belle:2019oag,BELLE:2019xld}, although with limited sensitivity. 
The measurements of $R_K$ are depicted in Fig.~\ref{fig:RK}.

Here and in the following charge-conjugated decay modes are implied. Experimentally, these ratios are constructed as either `simple' ratios of decays to different lepton flavors, or `double' ratios, with a further normalization to resonant decays. Both definitions usefully arrange efficiencies in ratios as well, so that many sources of systematic uncertainties cancel.

Because of the theoretical cleanliness of the underlying observables, the ensemble of these ratio measurements constitutes the centerpiece of three `$b \to s$ anomalies' --- the other two occurring in branching-ratio data to muons, and in angular analyses of differential decay data, that are discussed in a separate document~\cite{Altmannshofer:2022hfs}.

\begin{figure}[tb]
\centering
\includegraphics[width=0.7\textwidth]{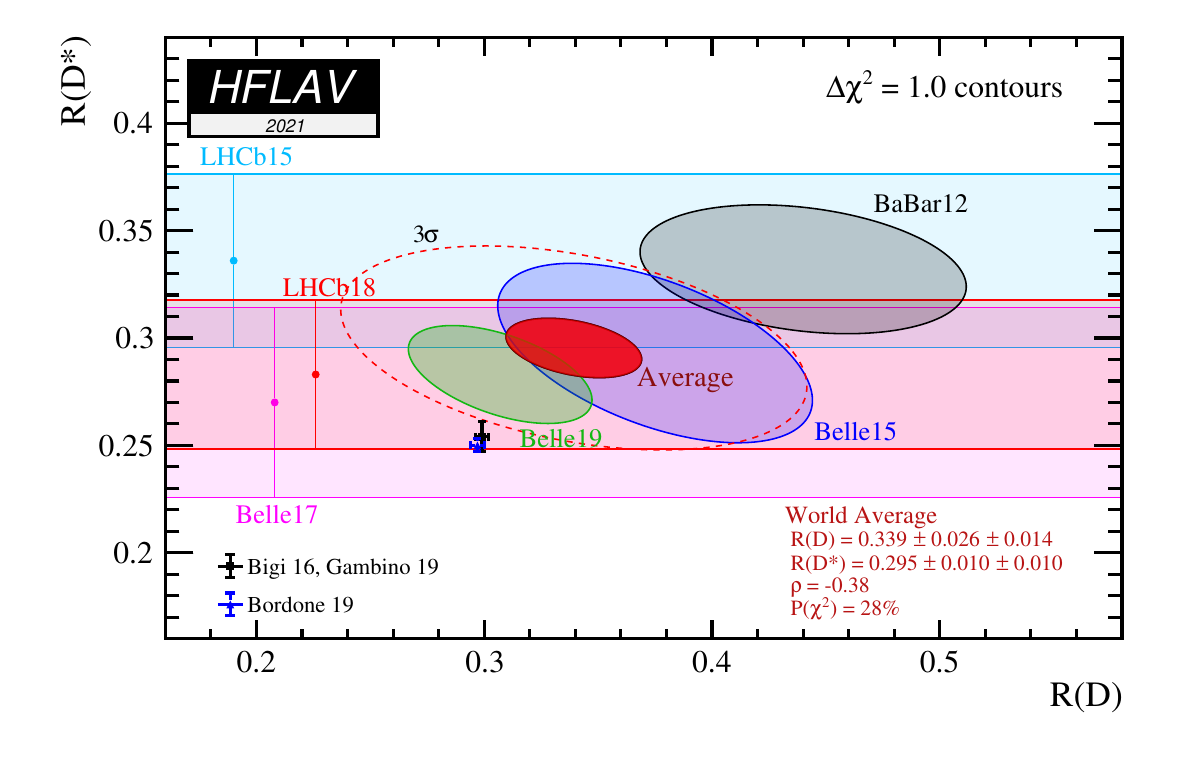}%
\caption{Latest experimental constraints on $R(D)$ and $R(\Dstar)$. Plot taken from the 2021 HFLAV web update (see also Ref.~\cite{Amhis:2022mac}).}\label{fig:RD}
\end{figure}

A second set of discrepant lepton-universality ratios concerns semileptonic \bquark $\to c$ decays:
\begin{equation}\label{eq:RD}
    R(\D^{(*)}) = \frac{%
            \BF(\decay{\Bb}{D^{(*)}\taum\neub})
          }{%
            \BF(\decay{\Bb}{D^{(*)}\ellm\neub})
            },
\end{equation}
with $\D^{(*)}$, \B any valid combination of charm and beauty hadrons, and \ellm a muon or an electron. Note the different notations in the $b \to s$ vs. $b \to c$ data, $R_X$ vs. $R(X)$. The present data shows a deviation of about three standard deviations from the SM expectation in the two-dimensional chart of $R(D)$ versus $R(\Dstar)$. Their combination is advantageous as experiments reach a better precision and control of (anti-)correlations when measuring both observables simultaneously. The deviation is mostly driven by a 2012 BaBar measurement~\cite{BaBar:2013mob}, see Fig.~\ref{fig:RD}. Updates from LHCb and Belle~II are eagerly awaited.

If lepton universality does not hold in decays of \bquark hadrons, it is natural to search for confirmation in yet unobserved processes, for example 
purely leptonic decays to two electrons, or to two tau leptons. The latter escape present sensitivities, similarly as the very constraining \decay{b}{s\neu\neub} decays.

Moreover, if one abandons lepton universality, there is no {\it a priori} reason to stick to lepton flavor conservation~\cite{Glashow:2014iga} --- which is known to be violated in neutrinos --- and searches for processes forbidden in the Standard Model become a high priority. 
So far no sign of charged-lepton flavor violation in \decay{B}{\ell\ell'} or \decay{b}{s\ell\ell'} processes has been seen yet and branching-ratio limits now reach the $10^{-9}$ range.

All the above applies to \bquark decays. Similar studies are needed in charm decays, that provide access to couplings to up-type quarks. Rare charm decays are notably more difficult than their beauty counterparts because the short-distance contributions are often subleading with respect to long-distance ones due to intermediate light resonances. In this discussion, ``short-distance'' loosely refers to the parts of the respective amplitudes that are perturbatively computable to any desired order in QCD $\times$ QED. The accuracy of the theory calculation is less of a strict requirement for null tests of the SM such as LFV processes, which are a clear sign of new physics the very moment they are measured to be non-zero.

\section{Experimental opportunities}
This section first presents the main experiments contributing to LUV and LFV measurements and then outlines the main processes of interest.

\subsection{Experiments}\label{sec:experiments}
The LHCb and Belle~II experiments are dedicated to measurements of rare processes with \bquark and \cquark hadrons and cover the whole programme described in this document. Many further experiments contribute in selected areas.
\subsubsection{LHCb}
LHCb is the LHC experiment optimised for flavor physics.
LHCb profits from the large \bquark\bquarkbar and \cquark\cquarkbar production cross-sections in $pp$ collisions in its forward acceptance:
$\sigma_{\bquark\bquarkbar}=144\pm21\mub$~\cite{LHCb-PAPER-2016-031} and $\sigma_{\cquark\cquarkbar}=2370\pm160\mub$~\cite{LHCb-PAPER-2015-041}.
All species of \bquark hadrons are produced, with typical ratios of 
4:4:2:1 for \Bp, \Bz, \Lb and \Bs hadrons, respectively. 
In addition, the \Bc meson is produced at a rate further suppressed by ${\cal O}(10^{-3})$~\cite{LHCb-PAPER-2019-033}.

Until now LHCb ran at a fixed instantaneous luminosity of $4\times10^{32}\cm^{-2}\sec^{-1}$, collecting data corresponding to
9\invfb. With its first upgrade being completed this year,
this luminosity is being increased by a factor five, to $2\times10^{33}\cm^{-2}\sec^{-1}$ for Runs 3 and 4 (covering the 2020ies)
with a target of 50\invfb~\cite{LHCb:Snowmass}. This is made possible by new detectors but also
by a full-software trigger that can cope with the high rates of 
\bquark and \cquark events at high efficiency. 
LHCb is planning a further upgrade~\cite{LHCb-TDR-023} which should lead to
a total integrated luminosity of 300\invfb. 

LHCb has excellent particle-identification and vertexing capabilities, which allow for good separation of signal and background. Requiring that the 
\bquark-hadron momentum point back to the primary $pp$ collisions sets stringent constraints on partially reconstructed backgrounds and often permits to close the kinematics even in the presence of undetected particles such as neutrinos. 

\subsubsection{Belle~II}
The Belle~II experiment is the successor of the Belle experiment which operated at the KEK-$B$ \epem collider in the 2000ies. It is a $4\pi$ magnetic spectrometer with subdetectors placed cylindrically around the beams.
Belle~II started operating in 2021, but still misses some elements of the pixel detector, to be installed in 2023~\cite{Forti:2022mti,BelleII:2022snowmass}.

Belle~II mostly runs at the \FourS resonance, which decays 
only to \B\Bb pairs. This results in comparatively clean events with ${\cal O}(10)$ final-state particles. While the Belle data set is just short of 1\invab, Belle~II targets 50\invab at an unprecedented instantaneous luminosity of $10^{35}\cm^{-2}\sec^{-1}$. 
These large luminosities are mitigated by a lower \B\Bb cross-section of about $1.1\nb$. 
Belle~II will also collect sizeable charm-hadron samples.

At the \FourS resonance the system of the two \B mesons has a known center-of-mass energy, which for fully reconstructed decay modes is used to identify signal peaks both in reconstructed mass and energy. For partially reconstructed decay modes this setup offers the possibility of fully reconstructing one (tag) \B meson, which determines the charge and four-momentum of the other (signal) \B meson. This feature can be used to search for
\decay{B}{\text{\emph{invisible}}} or to determine the momentum of the neutrino
in semileptonic \B decays, notably those involving \Ptau~leptons. 
Considerable effort is invested in optimising such 
event tagging methods; see {\eg} Refs.~\cite{Keck:2018lcd,Belle-II:2021rof}.

\subsubsection{LHCb and Belle~II compared}
The rule of thumb is that for \B-meson decays to charged particles, LHCb with 50\invfb will collect signal yields larger than those of Belle~II with 50\invab by a factor 5 to 10. When neutrals are involved the yields are more similar, while Belle~II will likely be superior for modes with neutrinos. 
LHCb will additionally collect heavier \bquark hadrons, such as \Bs, \Bc, and \Lb.

In general Belle~II and LHCb will be in competition for a limited number of measurements and otherwise be mostly complementary as they will shed light on the same parton-level physics processes from different angles.

\subsubsection{ATLAS and CMS}
The ATLAS and CMS experiments at the LHC also profit from large cross-sections and
run at a higher luminosity than LHCb. Their flavor physics capabilities are however limited compared to LHCb by much more stringent trigger requirements (notably on transverse momentum), a (presently) lower mass resolution, larger charged multiplicities, and very limited hadron-identification capabilities. They do however complement LHCb for decay processes with muons. They potentially can measure $R_X$ and $R(D^{(*)})$ ratios; however this is still to be demonstrated. It is therefore difficult to estimate their future contribution to these measurements.
No sensitivity projections to lepton-flavor or lepton-universality violation are given in Ref.~\cite{ATL-PHYS-PUB-2022-018}. 

\subsubsection{BES III and SCTF}
The BES III experiment in Beijing is also operating at a high-luminosity \epem 
collider, but at a lower energy than Belle~II. It is optimised for charmonium spectroscopy
and charm physics. A successor Super $\tau$-Charm Factory (STCF) is proposed in Ref.~\cite{Cheng:2022tog}. Its aim is to collect 1\invab at a collision energy of 3.773\gev.

\subsubsection{Future colliders}
New \epem colliders at the \Z-boson resonance, such as the FCC-$ee$ proposal~\cite{Bernardi:2022hny} or CEPC~\cite{CEPCPhysicsStudyGroup:2022uwl}, combine the advantages
of $pp$ and \FourS colliders. The \bquark\bquarkbar cross-section is larger
than at the \FourS resonance, all species of \bquark hadrons are produced, 
and \bquark\bquarkbar events remain relatively clean (but are not limited to just \bquark hadrons).

Both FCC-$ee$ and CEPC expect to collect ${\cal O}(10^{12})$ \decay{Z}{\bquark\bquarkbar} events, which corresponds to 20 times more \Bz and \Bp mesons than Belle~II, to which \Bs, \Bc and \Lb hadrons uncovered by Belle~II are to be added. A striking feature of FCC-$ee$ is an excellent mass resolution.

The muon collider and FCC-$hh$ projects are too far in the future to be discussed here. 

\subsection[Lepton universality in \decay{\bquark}{\squark\ellell}]{Lepton universality in \boldmath \decay{\bquark}{\squark\ellell}}\label{Sec:blls}
Given the present status of \B anomalies, it is of paramount importance to precisely study \decay{\bquark}{\squark\epem} and \decay{\bquark}{\squark\mumu} processes. These include not only partial branching fractions in the form of $R_X$ ratios, but also angular distributions. More details can be found in Ref.~\cite{Altmannshofer:2022hfs}.

\subsubsection[{$R_X$ ratios}]{\boldmath $R_X$ ratios}
\begin{figure}
\centering
\includegraphics[width=0.7\textwidth]{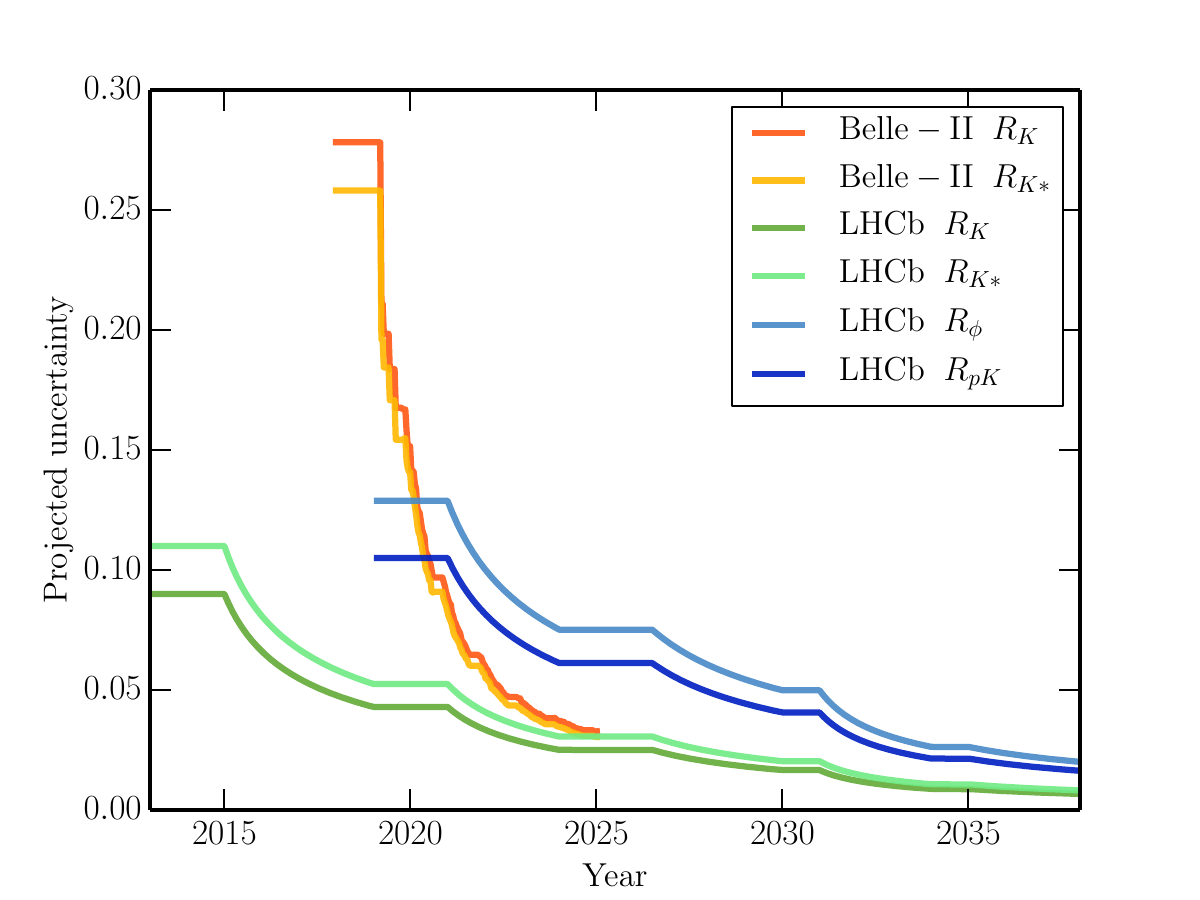}
\caption{Precision on $R_X$ ratios. Figure from Ref.~\cite{Bifani:2018zmi}. The time frame has slightly shifted since its publication.\label{fig:Future_RH}}
\end{figure}

The most accessible set of processes for $R_X$ ratios are \decay{\B}{\kaon\ellell}~\cite{LHCb:2021trn,BELLE:2019xld}, where \lhcb will have highest sensitivity to \decay{\Bp}{\Kp\ellell} while Belle~II performs similarly well for charged and neutral modes. Conversely, for \decay{\B}{\Kstar\ellell} LHCb will be most sensitive to the neutral mode (with \decay{\Kstarz}{\Kp\pim}).
While LHCb has measured LFU ratios with \KS~\cite{LHCb-PAPER-2021-038}, the yields are an order of magnitude lower than those with only charged final-state particles. Other decays will also be measured by LHCb, using \Bs and \Lb hadrons.
Prospects for future improvements in precision are shown in Fig.~\ref{fig:Future_RH}.

The challenges inherent in $R_X$-ratio measurements at hadron colliders deserve some more comments. These challenges are due to the already mentioned differences in electron vs.~muon efficiencies. One basic reconstruction challenge is consequence of electrons emitting much more bremsstrahlung than muons, whereby the reconstructed momentum is the momentum after emission, which differs from the momentum required for the dilepton invariant mass of the event. This problem has been the subject of constant scrutiny within the analyses. The numerous tests performed suggest that the effect is understood. Two representative examples of these tests include the following: {\em (1)} $R_K$ can be measured in the control region where the dilepton is emitted resonantly by the \jpsi or the \psitwos; {\em (2)} the electron efficiencies $\eps_e$ are calibrated in the \decay{\jpsi}{\epem} region and extrapolated to the signal region; the kinematic properties of electrons in these two regions are very similar.
An important outcome of these tests is that the ratio values obtained for electrons with kinematics in either of the resonant regions are well compatible with unity.

A further reassuring fact is that, although branching fractions to di-electrons generally come with inferior yields and larger systematic uncertainties with respect to branching fractions to di-muons, it is the latter that display discrepancies --- in particular the vast majority of $b \to s \mumu$ branching fractions (\eg~\cite{LHCb-PAPER-2014-006,LHCb-PAPER-2014-032,LHCb-PAPER-2015-023,LHCb-PAPER-2021-014}) --- whereas {\em all} branching fractions to di-electrons are SM-like within the quoted uncertainties. As a consequence, dismissing the $R_X$ measurements on the ground of unaccounted systematic effects in electrons is not straightforward --- how would such systematic uncertainties {\em not} manifest themselves in $\mc B(b \to s ee)$ data, that are SM-like, and instead result in $\mc B(b \to s \mumu)$ below the SM predictions, in basically all channels with large yields? In other words, ratios $\mc B(b \to s \mumu) / \mc B(b \to s e^+ e^-)$ below unity would be suspicious if the denominator were above the SM prediction, but instead it is the numerator which is below the SM, in $B \to K$ as well as in any other measured channels, including baryon modes.

\subsection[{\decay{\bquark}{\squark\taup\taum}}]{\boldmath\decay{\bquark}{\squark\taup\taum}}
Given the tensions in measurements comparing \decay{\bquark}{\squark\ellell} processes with muons and electrons, and those comparing light and \Ptau leptons in \decay{\bquark}{\cquark\ell\nu} it is crucial to measure \decay{\bquark}{\squark\taup\taum} processes to gain a complete picture.
Decays such as \decay{\B}{K^{(*)}\taup\taum} are however notoriously difficult to reconstruct. On top of the usual \decay{\bquark}{\squark\ellell} suppression they suffer from a reduced phase space which makes only the high-\qsq region above the \psitwos resonance accessible. In addition, the missing \Ptau neutrinos prevent a clean separation of signal from charmed backgrounds. The present limits are five orders of magnitude above the SM predictions and thus only sensitive to models predicting spectacular effects. Improvements will be made by Belle~II, who may reach the $10^{-4}$ range. To reach the SM signal, one may have to wait for FCC-$ee$ where, it is claimed, the excellent vertexing can reach a 5\% sensitivity on SM branching fractions~\cite{Bernardi:2022hny}.

\subsection[{\decay{\bquark}{\squark\neu\neub} decays}]{\boldmath\decay{\bquark}{\squark\neu\neub} decays}\label{Sec:bnns}
Rare decays involving neutrino pairs will be mostly Belle~II territory, thanks to their ability to fully analyse the event~\cite{Belle-II:2021rof}. Belle II expects uncertainties on the branching fraction of exclusive \decay{\B}{K^{(*)}\neu\neub} modes between 10 and 50\% of the SM expectation~\cite{BelleII:2022snowmass}. See Table~\ref{tab:exp} for details. First observations are thus expected in some of these processes. 

Further in the future, FCC-$ee$ or CEPC should be able to exploit their good vertexing resolution to measure decays not accessible to Belle~II with similar precision as above. The decays \decay{\Bs}{\phi\neu\neub} and \decay{\Lb}{\Lz\neu\neub} are most promising, and even \decay{\Bc}{\Ds\neu\neub} is feasible~\cite{Bernardi:2022hny,EF3_EF0-RF1_RF0-IF3_IF6-077}.

\subsection{Leptonic decays}\label{Sec:B2lnu}
Leptonic decays \decay{\Bp}{\ellp\neu_\ell} are accessible to Belle~II thanks to the possibility to reconstruct the rest of the event. While the present precision on the \decay{\Bp}{\mup\neu} and \decay{\Bp}{\taup\neu} decays are in the 20--50\% range, Belle~II expects 3\% with 50\invab. 

The decays \decay{\Bs}{\taup\taum}, \decay{\Bs}{\mumu} (see Ref.~\cite{Altmannshofer:2022hfs}) and \decay{\Bs}{\epem} will be dominated by LHCb (together with ATLAS/CMS for the dimuon mode). Limits on  \decay{\B}{\taup\taum} decays are now in the $10^{-3}$ range and are expected to reach $10^{-4}$ by the end of LHCb Upgrade II~\cite{LHCb-PUB-2018-009}. As the search is background-dominated, improvements scale as $\sqrt{\int {\cal L}dt}$, and thus are expectedly slow. To our knowledge no outlook is available for \decay{\Bs}{\epem}. It is expected that Belle~II will also set limits on the even more suppressed corresponding \Bz decays.

A particular type of $(\bar \bquark\cquark)(\neub \ell)$ coupling, related to those of  Sec.~\ref{Sec:clnu}, produces \decay{\Bc}{\ellp\neu} decays. FCC-$ee$ and CEPC expect a precision on the branching fraction of \decay{\Bc}{\taup\neu} below 1\%~\cite{Bernardi:2022hny,Zheng:2020ult}, and as many as $10^5$ \decay{\Bc}{\mup\neu} decays under SM assumptions~\cite{Bernardi:2022hny}.

\subsection[{Lepton universality in \decay{\bquark}{\cquark\ellm\neub}}]{Lepton universality in \boldmath \decay{\bquark}{\cquark\ellm\neub}}\label{Sec:clnu}

Decays involving \decay{\bquark}{\cquark\taum\neub_\tau} transitions are more challenging due to the multiple undetectable neutrinos in the final state. The \B factories have performed the most
precise measurements of $R(D)$ and $R(\Dstar)$ to date thanks to their ability to significantly
constrain the kinematics of these neutrinos by leveraging the knowledge of the \epem collision energy at these facilities~\cite{BaBar:2014omp}. Belle~II is expected to have the highest sensitivity to these measurements in the next decade.

In these decays an initial $b$-flavored hadron (a $\Bb$ meson or a $\Lb$ baryon) decays to a $c$-flavored one ($D^{(*)}$ or $\Lc$) plus a charged leptonic current $\ell^{\pm} \parenbar{\nu}$.
These decays occur in the SM already at tree level; they are namely not suppressed by a loop factor as is the case of the $b \to s$ semi-leptonic counterparts. Therefore these decays --- at least those to final-state muons and electrons --- have been used for a `NP-free' determination of the CKM entry $V_{cb}$. This assumes no NP-induced LUV in electron vs. muon modes~\cite{CLEO:2002fch,BaBar:2008zui,Belle:2015pkj,Belle:2017rcc,Belle:2018ezy}. The parameter $V_{cb}$ is often used as one of the four standard parameters that fully describe the CKM matrix --- for a discussion, see \eg Ref.~\cite{Buras:2020xsm}. At present, however, the ratios known as $R(D^{(*)}) \equiv \mc B(\Bb \to D^{(*)} \tau \nu) / \mc B(\Bb \to D^{(*)} \ell \nu)$ are in disagreement with the respective SM predictions at 3.3$\sigma$ as of the HFLAV June 2021 web update (see also Ref.~\cite{Amhis:2022mac}). As a consequence of these discrepancies, $R(D^{(*)})$, $V_{cb}$ and possibly other sensible parameters are now carefully fitted simultaneously, following various approaches discussed in Sec. \ref{sec:b_to_c_TH}. The discrepancies in $R(D^{(*)})$ constitute the fourth `$B$ anomaly' --- in addition to the three in $b \to s$ semi-leptonic transitions also mentioned in Sec. \ref{sec:most_promising}.

$R(D^{(*)})$ is best measured in events whose initial-state kinematics is known, as is the case at \B factories. This knowledge, as well as a large angular coverage, partly recovers the missing kinematic information due to the final-state neutrinos --- at least two. $R(D^{(*)})$ have been measured at BaBar and Belle in Refs.~\cite{BaBar:2012obs,Belle:2015qfa,Belle:2016dyj,Belle:2017ilt,Belle:2019rba,BaBar:2013mob}. LHCb has measured $R(D^*)$ in Refs.~\cite{LHCb-PAPER-2015-025,LHCb-PAPER-2017-017,LHCb-PAPER-2017-027}. Here the analysis strategy relies on first inferring the momentum of the parent $B$ meson from the flying direction estimated through the reconstructed decay vertex. Then, $R(D^*)$ is determined in a multi-dimensional fit, including different variables according to the decay modes --- hadronic or leptonic --- of the $\tau$ lepton. Similar strategies are used for the already mentioned $R(\jpsi)$~\cite{LHCb-PAPER-2017-035} and $R(\Lc)$~\cite{LHCb-PAPER-2021-044}. 

Knowledge of the $b \to c$ modes mentioned above, as well as additional ones, will steadily increase in the years to come, thanks to measurements at Belle~II~\cite{Belle-II:2018jsg} as well as LHCb~\cite{LHCb-PUB-2018-009}. Being Belle~II a lepton collider, it will have the same key advantages as discussed above --- and should even be sensitive to the $D^*$ and $\tau$ polarizations. Given the multiplicity of final states accessible, Belle~II could even be able to perform the inclusive measurements advocated in Refs.~\cite{Ligeti:2014kia,Mannel:2017jfk}. LHCb will also represent an important asset, for example because of the multiplicity of $R$-measurements it will be able to access --- including channels such as the $D_s$, the $\Lc$ and the $\jpsi$. One can expect the experimental precision of these measurements to be ultimately few percent.

\subsection{Leptonic universality in charm}\label{Sec:LUVcharm}
The STCF factory will allow a precision measurement of the ratios
\begin{equation}
    R_{D^+_{(s)}} = \frac{\Gamma(\decay{D^+_{(s)}}{\taup\neu_\tau})}{\Gamma(\decay{D^+_{(s)}}{\mup\neu_\mu})} \stackrel{\text{SM}}{=} \frac{m_\taup^2\left(1-\frac{m_\taup}{m_{D^+_{(s)}}}\right)^2}{m_\mup^2\left(1-\frac{m_\mup}{m_{D^+_{(s)}}}\right)^2},
\end{equation}
which are measured by BES III to 20\% and 5\% precision for \Dp and \Ds, respectively~\cite{Cheng:2022tog}. It is anticipated that this precision can be reduced to 0.4\% with STCF.

The ratio of \decay{\D}{\pi\mup\neu_\mu} to \decay{\D}{\pi\ep\neu_e} branching fractions is measured at BES III with about 4\% precision~\cite{BESIII:2018nzb}, which could also be significantly improved at STCF (and at Belle~II), though no precise extrapolation is available yet.

\subsection{Lepton flavor violation}\label{Sec:LFV}
Following from the above, searches for charged lepton flavor violating decays should be categorised into modes with \Ptau leptons, and modes without, i.e. with an electron and a muon.

The latter are relatively easy and build on the same experimental techniques as processes involving dielectrons and dimuons. The main difficulty is the absence of a control mode with the same final state (as \decay{\B}{\jpsi(\ellell)\Kp}), but that is hardly limiting for a new physics search. Limits at the few $10^{-9}$ level exist for processes as
\decay{\B}{\epm\mump}~\cite{LHCb-PAPER-2017-031}, or
\decay{\B}{\kaon\epm\mump}~\cite{LHCb-PAPER-2019-022},
and will be improved by Belle~II and LHCb. An improvement by an order of magnitude by the end of LHCb Upgrade II is reachable. Similar searches have been performed with charm~\cite{LHCb-PAPER-2020-007}. The larger cross-section may make charm a promising route to explore, although no sensitivity studies are available yet.

On the other hand, processes with \Ptau leptons as \decay{\B}{\taupm\mump}~\cite{LHCb-PAPER-2019-016} or
\decay{\B}{\kaon\taupm\mump}~\cite{BaBar:2012azg} suffer from much larger backgrounds. Innovative analysis techniques as the use of a $B_{s2}^{*0}$ tag~\cite{LHCb-PAPER-2019-043} may be required. The limits, presently in the few $10^{-5}$, may reach the $10^{-6}$ range.

\begin{table}[tb]\centering
\caption{Summary of expected experimental precision for selected observables in \bquark physics. Numbers from Refs.~\cite{LHCb:Snowmass} (LHCb) and~\cite{BelleII:2022snowmass} (Belle II; when available the improved scenario is taken). Here $\mu$ stands for the signal strength relative to the SM. More numbers are given in the text, notably for STCF and FCC-$ee$.}\label{tab:exp}
\begin{tabular}{L|CCCC}
\text{Observable} & \text{Current} & \text{LHCb U1} & \text{Belle II} & \text{LHCb UII} \\
\hline
R_\kaon ([1,6]\gevgevcccc) & 0.044\text{~\cite{LHCb-PAPER-2021-004}} & 0.025 & 0.036 & 0.007 \\
R_\Kstar ([1,6]\gevgevcccc) & 0.12\text{~\cite{LHCb-PAPER-2017-013}} & 0.031 & 0.034 & 0.009 \\
\hline
 R(\Dstar) & 0.014\text{~\cite{Amhis:2022mac}} & 0.007 & 0.003 & 0.002 \\
 R(\D)     & 0.030\text{~\cite{Amhis:2022mac}} &       & 0.004 \\
 R(\jpsi)  & 0.24\text{~\cite{LHCb-PAPER-2017-035}} & 0.07 & & 0.02 \\
\hline
\mu(\decay{\Bp}{\Kp\neu\neub}) & 0.7\text{~\cite{BaBar:2012azg}} && 0.08 & \\
\mu(\decay{\Bz}{\Kstarz\neu\neub}) & 1.8\text{~\cite{Belle:2017oht}} && 0.34 & \\
\hline
\BF(\decay{\B}{\Kstar\taup\taum}) & <2\times10^{-3}\text{~\cite{Belle:2021ndr}} & & <5.3\times10^{-4} & \\
\BF(\decay{\Bp}{\mup\neu}) & 50\%\text{~\cite{Belle:2019iji}} & & 2.5\% & \\
\BF(\decay{\Bp}{\taup\neu}) & 22\%\text{~\cite{PDG}} & & 3.0\% & \\

\end{tabular}
\end{table}
\section{Theory aspects and challenges}

\subsection[{Lepton Universality in Semi-Leptonic $b \to s$ Decays} ]{\boldmath Lepton Universality in Semi-Leptonic $b \to s$ Decays} \label{sec:LUV_TH}

Lepton-universality tests are usually constructed as ratios~\cite{Hiller:2003js,Wang:2003je}, or differences~\cite{Altmannshofer:2015mqa,Capdevila:2016ivx,Serra:2016ivr}, of two semi-leptonic  branching ratios, whereby the two concerned branching ratios differ only by the lepton flavor. In particular, since branching ratios are integrals over phase-space of their differential counterparts, the dilepton invariant mass range has to be the same between the two branching ratios concerned.

These ratios are by construction tests of a near-symmetry of the SM, lepton universality. Within the SM gauge interactions couple universally to matter, and the only non-universal dynamics arises from Yukawa interactions and is proportional to the mass of the concerned matter particle. Such effects have been addressed in ratios such as $R_X$ (Eq.~\ref{eq:RX})
and are minuscule~\cite{Bobeth:2007dw,Hiller:2003js}.\footnote{For small enough $q_{\rm min}^2$ one gets close to the lower endpoint in the muon channel and there is LUV by lack of phase space.} QED effects may also lead to lepton universality violation (LUV), in particular from collinear corrections (due to photons of {\em arbitrary} energy within the kinematic limit) $\propto \ale \log(m_\ell / \Delta)$, with $\Delta$ denoting any other scale in the problem. This may include inherently physical scales such as $m_B$ or $\LaQCD$, or scales induced by the definition of the observable, in particular $q_{\rm min, max}^2$. The above corrections have been evaluated analytically in Refs.~\cite{Bordone:2016gaq,Isidori:2020acz,Isidori:2022bzw} in the framework of a pointlike-meson Lagrangian, and the resulting spectrum has been compared to the one produced by the {\tt PHOTOS} simulation~\cite{Davidson:2010ew}.

$R_X$ ratios like in Eq.~(\ref{eq:RX}) are advantageous from both experimental and theoretical viewpoints. Form-factor induced uncertainties --- that usually constitute the largest source of uncertainty in branching ratios\footnote{It is understood that the phase-space integration is defined so as to exclude resonant regions.} --- cancel to a large extent in such ratios. 

Two main sorts of theoretical uncertainties can affect LUV observables. The first are QED corrections with large logs; the second may arise from the imperfect cancellation of hadronic uncertainties. This occurs as the prediction of LUV observables departs from the lepton-universal limit --- \eg as $R_{K^{(*)}}$ depart from unity. Let us first discuss QED corrections. Although these contributions are proportional to a small coupling, $\alpha / \pi \approx 2 \times 10^{-3}$, kinematic effects in $\Bb \to \Kb \ell^+ \ell^-$ can enhance them to ${\cal O}(\alpha / \pi) \log(m_\ell / m_B) \gtrsim 2\text{--}3\%$. A first analysis --- single-differential in the dilepton invariant mass squared $q^2$ --- was performed in Ref.~\cite{Bordone:2016gaq}. The authors find agreement with {\tt PHOTOS}~\cite{Davidson:2010ew} at per mil level, and assign $R_K$ an uncertainty of 1\%. A more general analysis~\cite{Isidori:2020acz} calculates the full matrix elements, \ie\ both the real and virtual components, and studies the fully differential decay rate. A further recent analysis~\cite{Isidori:2022bzw} constructs a dedicated simulation to describe QED corrections to $\Bb \to \Kb \ell^+ \ell^-$. Besides a comparison with {\tt PHOTOS}, this tool is also used to investigate effects of charmonium resonances. All of Refs.~\cite{Bordone:2016gaq,Isidori:2020acz,Isidori:2022bzw} adopt an EFT Lagrangian description, i.e scalar QED. The aim is to capture effects beyond collinear $\log(m_\ell / m_B)$ terms. Using arguments of gauge invariance, Ref.~\cite{Isidori:2020acz} also suggests that there are no leftover $\log(m_\ell / m_B)$ contributions due to structure dependence.

\begin{figure}[t]
  \begin{center}
    \includegraphics[width=\textwidth]{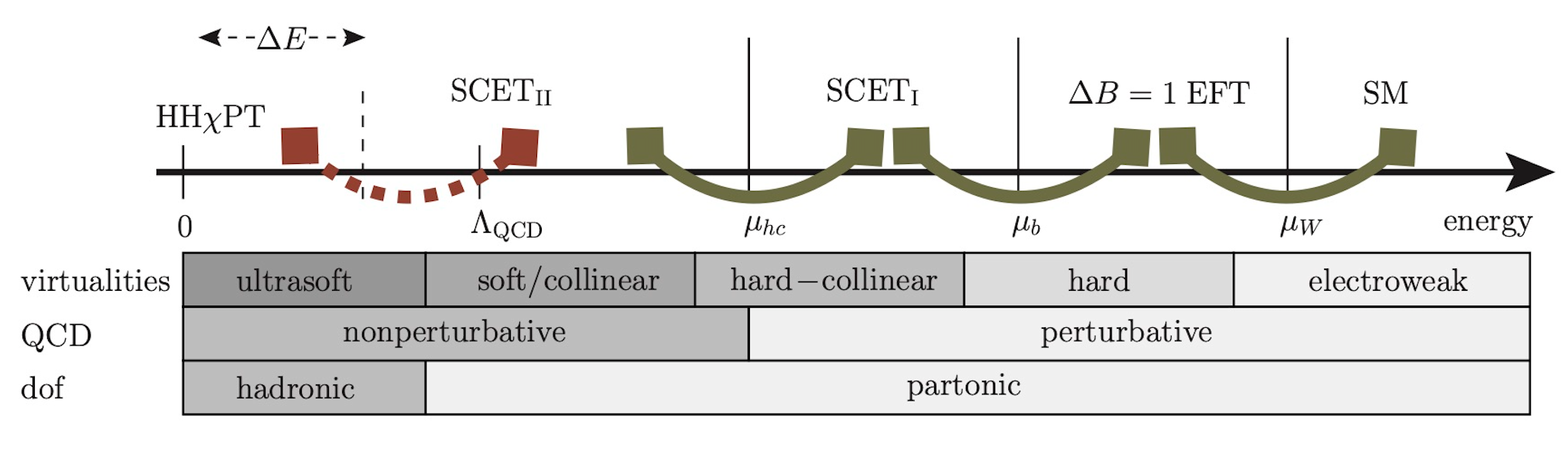}
  \end{center}
  \caption{Tower of theories and scales relevant for the description of semi-leptonic heavy-to-light lepton-universality ratios including QED corrections. $\Delta E$ denotes the minimum energy for single-photon detection. Figure taken from Ref.~\cite{szafron:CKM2021}.
\label{fig:szafron}
}
\end{figure}

The subject of QED corrections to $b \to s$ decays is clearly central to the theoretical control of LFU observables. This subject has a number of open challenges (see \eg Ref.~\cite{szafron:CKM2021}). On the sheer phenomenological side, one may expect that structure-dependent corrections for semi-leptonic heavy-to-light lepton-universality ratios --- including $R_{K^{(*)}}$ --- depart from unity by terms of ${\cal O}(\alpha) \times {\cal O}(\log(m_\mu^2 / \dots)$, where ellipses denote {\em any} scale in the experimental observable. Before any more refined argument, such terms may give a few-\% uncertainty at worst. One should also emphasize the complementarity of lattice evaluations of QED corrections for heavy mesons with respect to the EFT approach. In particular, lattice QCD can estimate $\log(m_B / \Lambda_{\rm QCD}) \sim 3$ terms, that are comparable in size with those captured within the EFT approach. Novel ideas and applications are being pursued, both at low and high dilepton $q^2$, corresponding to respectively large and small photon energies $E_\gamma$. Strictly speaking, only the low-$E_\gamma$ case is directly relevant to the discussion in the previous paragraph.\footnote{It should be noted however that inclusion of a hard photon lifts the chiral suppression in $\BdorBs \to \ellell$ decays. The electronic and muonic modes have thereby comparable rates. At facilities where electron and muon efficiencies are comparable (which in principle includes LHCb starting from Run 3), one could then consider $R_\gamma \equiv \int d {\BF}(\Bs\to \mumu \gamma) / \int d {\BF}(\Bs\to \epem\gamma)$ with an energetic photon~\cite{Guadagnoli:2017quo,Kozachuk:2017mdk}. LQCD calculations of $B \to \gamma$ form factors (FFs) for energetic photons would be crucial for such observables.} For large $E_\gamma$, it has been noted that the required correlator --- a weak and an electromagnetic current insertion between the external hadronic states --- has the desired behavior for large Euclidean time, if the matrix element is between a $B$ and the vacuum~\cite{Kane:2019jtj,Desiderio:2020oej}. This advantageous property holds specifically for radiative {\em leptonic} decays --- \ie\ it does not hold in the semi-leptonic case.\footnote{We thank Stefan Meinel for insightful conversations on this matter.} In addition, the $B \to \gamma$ FFs in this $q^2$ region have been calculated in several recent studies based on QCD factorization and soft collinear effective theory~\cite{Beneke:2020fot} or on light-cone sum rules~\cite{Wang:2016qii,Wang:2018wfj,Pullin:2021ebn,Janowski:2021yvz}. The factorization approach ensures that the main parametric dependence of the leading-power FFs in this region be on the $B$-meson light-cone distribution amplitude --- hence the challenge is to control this quantity to a satisfactory accuracy, see approaches in Refs.~\cite{Beneke:2011nf,Wang:2019msf}, and related discussions in Refs.~\cite{Zhao:2020bsx,Galda:2022dhp}. Even with an accurate determination of this quantity, the $B \to \gamma^*$ FFs in this region get formally next-to-leading-power --- but actually numerically dominant --- contributions from resonant regions that completely escape a factorization description~\cite{Beneke:2020fot}. For small $E_\gamma$, the main underlying problem is to define IR-safe LQCD quantities. A novel approach to this problem was put forward in Refs.~\cite{Giusti:2017dwk,DiCarlo:2019thl} and first applied to the $K_{\ell 2}$ case. In a nutshell, the idea is to use the continuum width, calculated within scalar QED, in order to cancel the IR divergences in the width from LQCD. This `subtraction' is performed for each photon momentum considered within the lattice simulation. The main challenge for this idea to fully capture QED logs non-perturbatively --- also in real-emission contributions --- is to go beyond the assumption of scalar QED, which implies a cutoff on $E_\gamma$ well below $\LaQCD$. However, a determination of $B \to \gamma^*$ FFs, even for $E_\gamma \sim$ GeV, seems within reach with this method.
From a theory point of view, the basic challenge is to perform a non-perturbative matching between the point-like EFT and the microscopic description. The corresponding sequence of theories to be matched to one another is well summarized in Fig.~\ref{fig:szafron}, taken from Ref.~\cite{szafron:CKM2021}. The necessity to carefully include {\em hard} but collinear photons has been elucidated in a benchmark application to $\Bs \to \mu^+ \mu^-$~\cite{Beneke:2017vpq,Beneke:2019slt}, which identifies single and double $\log(m_b \omega / m_\ell^2)$ terms ($\omega \approx \Lambda_{\rm QCD}$), that however largely compensate for seemingly accidental reasons.
Additional steps towards a systematic treatment of QED in charmless $\Bz \to \pi^+ \pi^-$ and in heavy-to-heavy decays have been undertaken in Refs.~\cite{Beneke:2020vnb,Beneke:2021jhp}.
It should also be noted that, as soon as non-perturbative soft matrix elements are evaluated within QCD$\times$QED, light-cone distribution amplitudes have to be generalized accordingly. This generalization is accomplished in Ref.~\cite{Beneke:2021pkl}.

The second source of theory uncertainty in ratio observables is the imperfect cancellation of the uncertainties induced by long-distance physics as the branching ratios in the numerator vs. the denominator of the ratio depart from the lepton-universal limit. Such departure may be induced by the different phase space available to the different lepton-flavor combinations in the numerator vs. the denominator --- as is the case \eg in $R_{K^*}$ for very low $q^2$~\cite{Bordone:2016gaq} or for $R(D^{(*)})$ --- or else it may due to LUV new physics. In this discussion we focus on the latter possibility, which has been explored in \eg Ref.~\cite{Altmannshofer:2021qrr}. This paper addresses the question of the validity of the approximation of evaluating the theory covariance matrix at the SM point. This approximation is expected to hold to the extent that NP is small with respect to the respective SM contribution. However, global analyses include observables whose theory uncertainty is negligible as compared to the experimental one only at the SM point. Examples include lepton-flavor universality tests like $R_{K^{(*)}}$, because the larger LUV NP contributions, the less efficient the cancellation of hadronic uncertainties between numerator and denominator. The correct procedure in this case is to re-evaluate the theory correlation matrix at each NP point being considered in the global fit. To have an idea of the possible impact of such uncertainty, let us consider $R_K$. The SM uncertainty is quoted as 1\%~\cite{Bordone:2016gaq,Isidori:2020acz}, which is small with respect to the current 5\% experimental uncertainty~\cite{LHCb:2021trn}. An ${\cal O}(15\%)$ LUV contribution from NP will multiply hadronic contributions known to, say, ${\cal O}(30\%)$ (from FFs {\em squared}). This translates into a contribution to the theory uncertainty of, again, 5\%, which is no more negligible with respect to the experimental uncertainty. A similar word of caution applies to other LFU ratios. An additional class of LFU tests are the quantities known as $D_{P_i^\prime} = Q_i$~\cite{Altmannshofer:2015mqa,Capdevila:2016ivx,Serra:2016ivr}. For them, current experimental uncertainties are completely dominant~\cite{Belle:2016fev} with respect to theory uncertainties, and accuracy projections suggest that the level of few percent~\cite{Albrecht:2017odf} is a longer-term prospective than in ratio tests. $Q_5$ may help distinguish genuinely LUV v.s lepton-universal NP contributions~\cite{Alguero:2018nvb}. A universal such shift to $C_9$ has a neat theoretical interpretation~\cite{Bobeth:2011st,Crivellin:2018yvo}, that naturally connects $b \to s \ell\ell$ and $b \to c \ell^- \bar{\nu}$ anomalies. This connection was found to work quantitatively in~\cite{Aebischer:2019mlg}.

\subsection[{Lepton Flavor Violation in $b \to s$ (Semi-)Leptonic Decays}]{\boldmath Lepton Flavor Violation in $b \to s$ (Semi-)Leptonic Decays} \label{sec:LFV_TH}

\noi The three `$B$ anomalies' hitherto discussed --- in $q^2$-integrates differential rates, in angular analyses, and in ratio tests --- coherently suggest that the discrepant measurements are all and only those to dimuons, while modes to dielectrons are SM-like within large uncertainties, and modes to ditaus are still too weakly constraining. These experimental facts suggest new physics hierarchically coupled to the different lepton generations~\cite{Glashow:2014iga}.

New LUV dynamics is generally accompanied by new lepton-flavor-violating (LFV) dynamics as well. Here, `generally' means that this is the expectation in the absence of further assumptions. In other words, in the same way as one could explain a diagonal CKM matrix --- had it been the case chosen by nature --- one can explain the size of LUV in $R_K$, and concurrently forbid non-standard (i.e. non-zero) LFV at the price of suitable assumptions --- whether dynamics or a symmetry mechanism for example. One consistent avenue to prevent measurable LFV in the presence of measurable LUV is to extend the peculiar lepton-flavor symmetries of the SM to hold also for the NP dynamics, see \eg Refs.~\cite{Celis:2015ara,Alonso:2015sja}.

It is clear that the above question can only be settled by experiment --- but it is a crucial question on the structure of the putative new dynamics. For guidance, one may ask oneself what is the general size of the expected LFV effects in the presence of LUV as large as observed~\cite{Glashow:2014iga}. The latter are of ${\cal O}(15\%)$ according to data, as opposed to unobservable effects within the SM. By a general argument~\cite{Glashow:2014iga} this figure suggests likewise measurable LFV effects (see also Refs.~\cite{Guadagnoli:2015nra,Guadagnoli:2016erb}). The starting point is the observation that all $b \to s$ data are explained at one stroke by a 4-fermion operator composed of a left-handed (LH) quark times a left-handed lepton structure~\cite{Hiller:2014yaa}. Importantly, the latter comes with a lepton-generation dependent Wilson coefficient, because $\mumu$ data hint at new effects, whereas \epem data do not. Such pattern suggests a purely $3^{\RD}$-generation LH $\times$ LH interaction, as the result of integrating out new states at some scale above the EWSB scale. As a consequence, the fields in this interaction are generally not mass eigenstates. The unitary transformations required to express the interaction in this basis will extend this interaction to generations other than the heaviest, and generally yield LUV along with LFV. The two sets of effects may be parametrically related to each other, because LUV is measured (through $R_K$). One thereby obtains ballpark estimates for LFV decays, such as $B \to K \tau \mu$, with branching ratios around $10^{-8}$~\cite{Glashow:2014iga}. In a nutshell, this order of magnitude arises as the product of $\mc B(B \to K \mumu) \approx 4 \times 10^{-7}$, times the departure of $R_K$ from unity, squared --- which yields a number around $10^{-8}$ --- times {\em ratios} of products of the above mentioned unitary rotations that lead to the mass eigenbasis. Since not all these ratios can be much smaller than unity (even if all unitary mixings are small numbers, there is no reason why also the ratios of such mixings should all be small numbers), one may expect that some LFV decay branching ratio be in the nominal ballpark of $10^{-8}$~\cite{Glashow:2014iga}.

The above picture must withstand certain constraints. In particular, the mentioned LH $\times$ LH interaction arising above the EWSB scale, it should be made compliant with $SU(2)_L$ symmetry~\cite{Bhattacharya:2014wla}. By closing the quark loop and connecting it to a further lepton pair through a gauge boson, this interaction then yields LFV effects in lepton decays, for example $\tau \to 3 \mu$~\cite{Feruglio:2017rjo,Feruglio:2016gvd}.

In recent years, the subject of LFV in semi-leptonic $B$ decays --- and even its possible connections with other flavored sectors --- has been extensively explored from a phenomenological point of view, see  Refs.~\cite{Crivellin:2015era,Hiller:2016kry,Becirevic:2016oho,Kumar:2016omp,Crivellin:2016vjc,Bordone:2017lsy,Hazard:2017udp,DAmbrosio:2017wis,Borsato:2018tcz,Mandal:2019gff,Gherardi:2019zil,Descotes-Genon:2020buf,Marzocca:2021miv,Becirevic:2016zri}. Expressions for the full angular distributions of the $B \to K^{(*)} \ell_1 \ell_2$ have been discussed in Ref.~\cite{Becirevic:2016zri}.
Many scenarios predicting LFV signals have been advocated, all the way from EFT approaches, to simplified gauge or LQ models, or composite Higgs sectors, or UV-complete models~\cite{Bhattacharya:2014wla,Crivellin:2015era,Gripaios:2014tna,Gripaios:2015gra,Calibbi:2015kma,Greljo:2015mma,Sahoo:2015pzk,Lee:2015qra,Boucenna:2015raa,Fajfer:2015ycq,deMedeirosVarzielas:2015yxm,Falkowski:2015zwa,Guadagnoli:2016erb,Duraisamy:2016gsd,Feruglio:2016gvd,Hiller:2016kry,Bhattacharya:2016mcc,Becirevic:2016oho,Chiang:2016qov,Boucenna:2016qad,Bordone:2017anc,Choudhury:2017ijp,Becirevic:2017jtw,Choudhury:2017qyt,Becirevic:2018afm,Hati:2018fzc,Rocha-Moran:2018jzu,Guadagnoli:2018ojc,Bordone:2018nbg,Kumar:2018kmr,Sheng:2018qtp,Bhattacharya:2018ryy,Angelescu:2018tyl,Sheng:2018ylg,Cornella:2019hct,Das:2019omf,Hati:2019ufv,Hati:2020cyn,Angelescu:2020uug,Greljo:2021xmg,Becirevic:2022tsj}.
All the modes discussed are realistically within reach at present facilities and/or at their upgrades. As a matter of fact, a detailed program of experimental searches has blossomed. Recent searches at LHCb include Refs.~\cite{LHCb-PAPER-2019-016,LHCb-PAPER-2019-043}, with more ongoing. For further details, see Sec. \ref{Sec:LFV}.

\subsection[{Semi-Leptonic $b \to s$ Modes with $\tau$ leptons}]{\boldmath Semi-Leptonic $b \to s$ Modes with $\tau$ leptons}

\noi To reiterate, the pattern of semi-leptonic $b \to s$ decay data --- SM-like in modes to electrons, discrepant to ${\cal O}(10\%)$ in modes to muons, still insensitive to the SM signal in modes to taus --- suggests new physics hierarchically coupled to the generations of matter~\cite{Glashow:2014iga}. It is worth noting that, for the third generation of leptons, this conclusion relies on the limited knowledge of the relevant modes: $\mc B(B^+ \to K^+ \tau^+ \tau^-)$ and $\mc B(B^+ \to K^+ \tau^\pm \mu^\mp)$ set weakly constraining bounds of ${\cal O}(10^{-3})$ and ${\cal O}(10^{-5})$ respectively, and hence order-of-magnitude signals from new physics are possible. In these circumstances, the above modes, as well as $\Bs \to \taup\taum$, are the perhaps most crucial test of the overall theory understanding~\cite{Glashow:2014iga}. More general surveys have been performed, including in particular $B \to K^{(*)} \taup\taum$ and $\Bs\to \phi \taup\taum$. With minimal assumptions and an EFT approach, one generally expects large enhancements of around three orders of magnitude with respect to the SM~\cite{Capdevila:2017iqn}. Several works mentioned in Secs. \ref{sec:LFV_TH} and \ref{sec:model_building} also quote similar enhancements in the context of models.

\subsection[{Semi-Leptonic $b \to c$ modes}]{\boldmath Semi-Leptonic $b \to c$ modes}\label{sec:b_to_c_TH}

The departures of the measured $R(D^{(*)})$ from the SM predictions may be interpreted in terms of new effects in $\mc B(\Bb \to D^{(*)} \tau \nu)$, namely in the numerator branching ratio. In fact, if new physics is present in $b \to s \ellell$ transitions, and is caused by dynamics occurring above the EWSB scale, the new effects should, to some extent, `spill over' to $b \to c \tau \nu$ transitions as well, especially if one starts with the assumption that the new interaction is dominantly coupled to the 3$\RD$ generation~\cite{Glashow:2014iga}. In fact, the $b \to s$ and $b \to c$ anomalies are closely related by $SU(2)_L$ symmetry~\cite{Bhattacharya:2014wla,Alonso:2015sja}.

Similarly as $b \to s \ellell$ transitions, one important aspect of $b \to c \ell \nu$ processes is the theoretical control of FFs, that are functions of the leptonic invariant mass squared $q^2$, and that arise from the matrix elements of the concerned quark bilinear between external hadronic states. Their determination follows different approaches.
A first one is lattice QCD (LQCD), and is best suited for high $q^2$, possibly close to the upper endpoint $q^2_{\rm max}$. Results exist for several meson transitions, including $\Bb \to D^{(*)}$~\cite{Na:2015kha,MILC:2015uhg,FermilabLattice:2014ysv,Harrison:2017fmw,FermilabLattice:2021cdg} and, in the full kinematic range, $\Bs\to D_s^{(*)}$~\cite{McLean:2019qcx,Harrison:2017fmw,McLean:2019sds,Harrison:2021tol} as well as $\Bc \to \jpsi$~\cite{Harrison:2020gvo,Harrison:2020nrv}. Form-factor calculations, as well as phenomenological applications, exist also for $b \to c$ decays involving baryons, in particular $\Lz_\bquark \to \Lz_\cquark^{(*)}$~\cite{Detmold:2015aaa,Datta:2017aue,Meinel:2021rbm,Meinel:2021mdj}.
As concerns $\Bc \to \jpsi$ and the ensuing $R(\jpsi)$, it is quite intriguing that the precise SM prediction leads to a discrepancy with the experimental result that is compatible in magnitude and sign with the $R(D^{(*)})$ anomaly.
Besides the specific calculations above, the interested reader is also referred to the dedicated Snowmass 2022 White Paper~\cite{Boyle:2022uba}, as well as the comprehensive FLAG review~\cite{Aoki:2021kgd}.
A further approach is a QCD-inspired method known as QCD sum rules~\cite{Shifman:1978bx,Shifman:1978by} (see also Refs.~\cite{Balitsky:1988tpa,Braun:1997kw}, and for a modern viewpoint~\cite{Colangelo:2000dp}). In this case, light-cone sum-rule calculations of $\Bb \to D^{(*)}$ FFs~\cite{Gao:2021sav,Gubernari:2018wyi,Wang:2017jow,Faller:2008tr} at $q^2$ values $\le 5$ GeV$^2$ are extrapolated to large $q^2$ following different approaches (see below).

The decay $\Bb \to D \ell \nu$ is parameterized in terms of two FFs, often chosen to be the vector and the so-called scalar one. The FF dependence for $\Bb \to D^* \ell \nu$ is more complex --- although, as discussed in Ref.~\cite{Kim:2016yth}, the $R(D^*)$ prediction may be more robust than is $R(D)$'s. The reason is that the scalar FF contributes sizeably to the $\tau$ mode --- the numerator  in $R(D)$ --- whereas its contribution is negligible for light leptons (in both cases the reference is the vector FF contribution). In particular, better agreement of $R(D)$ with experiment would be possible if the scalar FF departed with respect to the current lattice evaluation~\cite{Na:2015kha,MILC:2015uhg} by $O(10\%)$ in some $q^2$ range below $q^2_{\rm max}$~\cite{Kim:2016yth}.

Starting from the calculations mentioned above, extrapolations are typically required to estimate the FFs in the full kinematic range required for $R(D^{(*)})$ and other observables. Several approaches exist for such extrapolations. A first one is due to Boyd, Grinstein and Lebed (BGL)~\cite{Boyd:1997kz}. One starts~\cite{Boyd:1994tt,Boyd:1995cf,Caprini:1995ha,Boyd:1995sq,Boyd:1995tg,Caprini:1995wq,Lellouch:1995yv,Caprini:1996fn,Becirevic:1996ki,Becirevic:1996vk,Caprini:1996vw,Boyd:1997qw,Caprini:1994fh} from the FF normalization in the heavy-quark-symmetric limit (\cite{Isgur:1989vq,Isgur:1990yhj,Shifman:1987rj,Eichten:1989zv}, for reviews see Refs.~\cite{Georgi:1991mr,Neubert:1993mb}), and the FF shape, as functions of the momentum transfer, are subsequently constrained by means of dispersion relations. These relate an {\em inclusive}-production rate with a two-point function that can be calculated perturbatively in QCD. BGL showed that inclusion of higher states in the sum over channels through which the inclusive rate is estimated significantly improves the shape constraints for $b \to c$ transitions. Global analyses~\cite{Bigi:2017jbd} exploit the constraining power of such relations. The form-factor parametrization can be further constrained through fundamental QFT requirements such as unitarity, analyticity and perturbative-QCD scaling as in Bourrely, Caprini and Lellouch (BCL)~\cite{Bourrely:2008za}. This approach aims at making the parameterization as model-independent as possible, while also avoiding explicit expansions in $\alpha_s$ or in inverse powers of the heavy-quark mass. Applications~\cite{Bigi:2016mdz,Bigi:2017njr,Jaiswal:2017rve} include a simultaneous determination of the $R(D^{(*)})$ ratios and of $V_{cb}$, that allows to also get insights on the ``exclusive vs. inclusive'' tension in this CKM entry.

An additional, somewhat separate approach starts again from the heavy-quark-symmetry FFs, but then focuses on the systematic inclusion of QCD corrections as well as power-suppressed --- both in $1/m_b$ {\em and} in $1/m_c$~\cite{Bernlochner:2017jka,Bernlochner:2018kxh,Bernlochner:2018bfn,Bordone:2019vic,Bordone:2019guc}. This method has also been applied to the simultaneous determination of $R(D^{(*)})$ ratios and of $V_{cb}$~\cite{Bernlochner:2017jka,Bernlochner:2018kxh,Bernlochner:2018bfn,Bernlochner:2022ywh}. Refs.~\cite{Bordone:2019vic,Bordone:2019guc} focus instead on the convergence of the power expansion --- they include $1/m_c^2$ corrections --- and on maximally constraining the FFs through calculations within light-cone sum rules, plus additional constraints from LQCD, QCD three-point sum rules and unitarity.

Finally, yet another approach to $V_{cb}$ and $R(D^{(*)})$ is the so-called ``dispersive method'', very recently put forward in Refs.~\cite{DiCarlo:2021dzg,Martinelli:2021frl,Martinelli:2021onb,Martinelli:2021myh,Martinelli:2022vvh}. This model-independent method was originally introduced for lattice calculations in Ref.~\cite{Lellouch:1995yv}. Within this approach, FFs are described without  assumptions on their functional dependence on the momentum transfer. By enforcing the dispersive bounds due to unitarity and analyticity, as well as the existing lattice-QCD data on FFs --- available at large momentum transfer only --- one determines the FFs in a model-independent way in the full kinematical range. This leads to the predictions $R(D) = 0.296(8)$ and $R(D^*) = 0.275(8)$, whose agreement with the measurements' world average is at the 1.3$\sigma$ level.

The possibility to enhance lepton-universality tests in the $b \to c$ sector through additional observables with the same underlying current, including leptonic decays, specific angular distributions, measurements sensitive to specific polarization fractions, high-$\pt$ signatures has been discussed in Refs.~\cite{Roy:2017dum,Colangelo:2018cnj,Rui:2018kqr,Asadi:2018sym,Greljo:2018tzh,Cohen:2018vhw,Boer:2018vpx,Blanke:2018yud,Dutta:2018zqp,Becirevic:2019tpx,Mandal:2019vwq,Isidori:2020eyd,Asadi:2020fdo,Ivanov:2020iad,Bhattacharya:2020lfm,Penalva:2022vxy}. 
Lepton-universality tests in $b \to c$ semi-leptonic in the baryonic sector are discussed in Refs.~\cite{Datta:2017aue,Boer:2018vpx,Hu:2020axt,Duan:2022uzm,Bernlochner:2022hyz}.

\subsection{Model-building considerations} \label{sec:model_building}

The interpretation of ${\cal O}(10-20\%)$ LUV effects in semi-leptonic decays has to face well-defined challenges. The `minimal' requirements that data seem to convey include the following: {\em (i)} the new dynamics explaining the $b \to s$ measurements must, directly or indirectly, involve the second and the third generation of quarks and leptons; {\em (ii)} it must yield large enough effects in the product of a quark times a charged-lepton bilinear, $J_q \times J_\ell$, and small enough effects elsewhere, in particular in flavor-changing $J_q \times J_q$ and $J_\ell \times J_\ell$ amplitudes. These requirements have `genetically' selected LQs as the preferred candidate for a dynamical explanation. In fact, requirement {\em (ii)} holds automatically, because $J_q \times J_\ell$ can occur at tree level, whereas $J_q \times J_q$ and $J_\ell \times J_\ell$ are automatically loop-suppressed --- at least for `genuine' LQs~\cite{Dorsner:2016wpm,Davidson:1993qk}.

More formidable challenges arise if one wants to explain $b \to s$ {\em and} $b \to c$ hints of LUV concurrently. At face value, i.e. to the extent that both sets of `anomalies' have comparable significances, it looks justified to take both datasets on an equal footing. There are also theory considerations supporting such a stance, in particular the fact that the two sets of anomalies convey the same underlying piece of information --- sizeable LUV --- in currents that, above the EWSB scale, are related by the SM $SU(2)_L$ symmetry --- as expected of new short-distance effects~\cite{Bhattacharya:2014wla,Alonso:2014csa,Buttazzo:2017ixm}. The problem with a simultaneous explanation is at the {\em quantitative} level, as $b \to s$ data hint at $\sim 10-20$\% shifts in a SM loop amplitude, whereas $b \to c$ data hint at comparably large shifts, but in a tree amplitude. Then, if one introduces a common effective, $SU(2)_L$-invariant structure to account for both shifts, a mechanism must also be supplied for the flavor-dependent coupling to produce more suppressed effects in $b \to s$ than in $b \to c$ --- the relative suppression being approximately a loop factor. Various such mechanisms have actually been proposed in the literature. One instance the LQ model in Ref.~\cite{Bauer:2015knc}, giving rise to tree- vs. loop-suppressed amplitudes (on the phenomenology of this model, see also Refs.~\cite{Becirevic:2016oho,Cai:2017wry}). Another instance is~\cite{Barbieri:2015yvd}, which proposes a minimally-broken $U(2)^5$ global symmetry~\cite{Barbieri:2011ci,Barbieri:2012uh}. In this case, $b \to c \tau \nu$ and $b \to s \mumu$ effects arise as respectively first and third order in the breaking parameter~\cite{Barbieri:2015yvd} (see also Refs.~\cite{Buttazzo:2017ixm,Greljo:2015mma,Bordone:2017anc}).

One further important aspect to face is that, for the model to be thoroughly testable, all processes relevant as constraints should be, at worst, log-dependent on the UV scale. As well known, this is not an issue for massive-scalar extensions. A handful of combinations of scalars are most popular as combined explanations of $b \to s$ and $b \to c$ LUV, namely the two scalar leptoquarks $S_1$ and $S_3$ --- see Refs.~\cite{Crivellin:2017zlb,Buttazzo:2017ixm,Marzocca:2018wcf}, and~\cite{Dorsner:2016wpm} for nomenclature; $R_2$ and $S_3$ --- see Refs.~\cite{Becirevic:2018afm,Becirevic:2022tsj}; the $S_1$ plus a charged singlet $\phi^+$~\cite{Marzocca:2021azj}.
However, massive new vectors do pose a problem, as certain (constraining) processes display power-like dependence on the UV scale --- for a reference discussion on this point see Ref.~\cite{Barbieri:2015yvd}. This problem applies to the vector leptoquark $U_1\sim ({\mathbf 3},{\mathbf 1})_{2/3}$ ~\cite{Alonso:2015sja,Calibbi:2015kma,DiLuzio:2017vat,Assad:2017iib,Calibbi:2017qbu,Bordone:2017bld,Barbieri:2017tuq,Greljo:2018tuh,Blanke:2018sro,Fornal:2018dqn,Heeck:2018ntp,Fuentes-Martin:2020hvc}, the most popular single mediator capable of explaining the $R_{K^{(*)}}$ and $R_{D^{(*)}}$ anomalies (see \eg Refs.~\cite{Barbieri:2015yvd,Hiller:2016kry,Bhattacharya:2016mcc,Buttazzo:2017ixm,Angelescu:2018tyl,Kumar:2018kmr}).\footnote{The option of $R_2$ as a single mediator was discussed in Ref.~\cite{Popov:2019tyc}.}

In great synthesis, the $U_1$ LQ may be UV-completed via an appropriate gauge group, such as the one in the Pati-Salam (PS) model~\cite{Pati:1974yy} --- a leptoquark vector mediator is the natural mediator between a quark and a lepton if lepton number is the fourth color. However, the PS group in its original version is not an option in the light of high-$\pt$ constraints, as spelled out in~\cite{DiLuzio:2017vat}, which require to separate the $SU(4)$ group from $SU(3)_c$, and to  enforce $g_4 \gg g_1$, $g_3$. Minimality seems then to point to an $SU(4) \times SU(3)^\prime \times SU(2)_L \times U(1)^\prime$, or 4321 model~\cite{DiLuzio:2017vat} (see also Refs.~\cite{Georgi:2016xhm,Bansal:2018eha}). Besides calculability (in principle at least) of the processes that in a simplified-$U_1$ approach would be power-divergent, this UV-complete construction allows to include in the picture one additional important insight: the $U_1$ does not come alone as a mediator. The 4321 model implies a $Z^\prime$ and a ``coloron'' mediator, and the signals --- \eg at colliders --- of this extended sector have to be studied jointly~\cite{Cornella:2021sby,Baker:2019sli,Faroughy:2016osc} and a generic expectation are excesses in di-tau tails. Besides the collider aspect, this scenario has also well-defined low-energy signatures and null tests. Interestingly, the model has by construction no tree-level contributions to the otherwise very constraining $B \to K^{(*)} \nu \bar \nu$ processes~\cite{Buras:2014fpa}; it predicts large signals in $b \to s \tau^+ \tau^-$ and $b \to s \tau \mu$ currents and in $\tau$ decays~\cite{DiLuzio:2018zxy,Cornella:2018tfd,Cornella:2019hct,Cornella:2021sby}.
A recent comparison between the different LQ scenarios vs. existing data can be found in Ref.~\cite{Angelescu:2021lln}.

Finally, additional UV-complete proposals --- either non-LQ models, or alternative mechanisms to generate a mass for {\em vectors}, or bosonic-mediator combinations other than those detailed above --- aimed at addressing both $b \to s$ and $b \to c$ anomalies include~\cite{Matsuzaki:2018jui,Kumar:2018kmr,Azatov:2018kzb,Faber:2018qon,Li:2018rax,Trifinopoulos:2018rna,Blanke:2018sro,DaRold:2019fiw,Balaji:2019kwe,Marzo:2019ldg,Trifinopoulos:2019lyo,DelleRose:2019ukt,Altmannshofer:2020axr,Saad:2020ihm,Saad:2020ucl,Hu:2020yvs,BhupalDev:2020zcy,BhupalDev:2021ipu,Crivellin:2022mff,Fuentes-Martin:2022xnb}.

\section*{Acknowledgments}
We warmly thank Gudrun Hiller for initial involvement in this write-up, and Stefan Meinel and Yu-Ming Wang for insightful comments on the manuscript.

\addcontentsline{toc}{section}{References}
\bibliographystyle{LHCb}
\bibliography{references,LHCb-PAPER,LHCb-TDR}

\end{document}